\documentclass[a4paper,12pt]{article}
\pdfoutput=1
\usepackage{graphicx, rotating}
\usepackage{hyperref}\usepackage{slashed}
\usepackage{ifpdf}
\ifx\pdfoutput\undefined
\usepackage[dvips,bookmarks]{hyperref}	% This is for arXiv.org
\else
\usepackage{hyperref}	% This is for pdftex
\fi
\hypersetup{colorlinks,bookmarksopen,bookmarksnumbered,citecolor=verdes,
linkcolor=blus,pdfstartview=FitH,urlcolor=rossos}
\def\hhref#1{\href{http://arxiv.org/abs/#1}{#1}} % in bibliography
\newcommand{\cm}{\,{\rm cm}}

\usepackage{multicol}
\usepackage{color}
\definecolor{rosso}{cmyk}{0,1,1,0.4}
\definecolor{rossos}{cmyk}{0,1,1,0.55}
\definecolor{rossoc}{cmyk}{0,1,1,0.2}
\definecolor{blu}{cmyk}{1,1,0,0.3}
\definecolor{blus}{cmyk}{1,1,0,0.6}
\definecolor{bluc}{cmyk}{1,1,0,0.1}
\definecolor{verde}{cmyk}{0.92,0,0.59,0.25}
\definecolor{verdec}{cmyk}{0.92,0,0.59,0.15}
\definecolor{verdes}{cmyk}{0.92,0,0.59,0.4}

\font\tenrsfs=rsfs10 at 12pt
\font\sevenrsfs=rsfs7
\font\fiversfs=rsfs5
\newfam\rsfsfam
\textfont\rsfsfam=\tenrsfs
\scriptfont\rsfsfam=\sevenrsfs
\scriptscriptfont\rsfsfam=\fiversfs
\def\mathscr#1{{\fam\rsfsfam\relax#1}}

\newcommand{\fig}[1]{~\ref{fig:#1}}
\newcommand{\eq}[1]{~{\rm (\ref{eq:#1})}}

\newcommand{\GeV}{\,{\rm GeV}}
\newcommand{\TeV}{\,{\rm TeV}}

\def\circa#1{\,\raise.3ex\hbox{$#1$\kern-.75em\lower1ex\hbox{$\sim$}}\,}

\newcommand{\DM}{{\rm DM}}

\newcommand{\beq}{\begin{equation}}
\newcommand{\eeq}{\end{equation}}

\def\circa#1{\,\raise.3ex\hbox{$#1$\kern-.75em\lower1ex\hbox{$\sim$}}\,}
\makeatletter

\def\art{\@ifnextchar[{\eart}{\oart}}
\def\eart[#1]#2#3#4#5#6{{\rm #2}, {#3 #4} {\rm (#6) #5} [arXiv:\-{\hhref{#1}}]}
\def\hepart[#1]#2{{\rm #2, arXiv:\-\hhref{#1}}}
\newcommand{\oart}[5]{{\rm #1}, {#2 #3} {\rm (#5) #4}}

\newcounter{alphaequation}[equation]
\def\thealphaequation{\theequation\hbox to
0.6em{\hfil\alph{alphaequation}\hfil}}
\def\eqnsystem#1{
\def\@eqnnum{{\rm (\thealphaequation)}}
\def\@@eqncr{\let\@tempa\relax \ifcase\@eqcnt \def\@tempa{& & &} \or
\def\@tempa{& &}\or \def\@tempa{&}\fi\@tempa
\if@eqnsw\@eqnnum\refstepcounter{alphaequation}\fi
\global\@eqnswtrue\global\@eqcnt=0\cr}
\refstepcounter{equation} \let\@currentlabel\theequation \def\@tempb{#1}
\ifx\@tempb\empty\else\label{#1}\fi
\refstepcounter{alphaequation}
\let\@currentlabel\thealphaequation
\global\@eqnswtrue\global\@eqcnt=0 \tabskip\@centering\let\\=\@eqncr
$$\halign to \displaywidth\bgroup \@eqnsel\hskip\@centering
$\displaystyle\tabskip\z@{##}$&\global\@eqcnt\@ne
\hskip2\arraycolsep\hfil${##}$\hfil& \global\@eqcnt\tw@\hskip2\arraycolsep
$\displaystyle\tabskip\z@{##}$\hfil
\tabskip\@centering&\llap{##}\tabskip\z@\cr}
\def\endeqnsystem{\@@eqncr\egroup$$\global\@ignoretrue} \makeatother

\begin{document}

\begin{center}
IFUP-TH/2008-36\hfill SACLAY--T08/184

\bigskip

{\huge\bf\color{magenta}
Gamma-ray and radio tests of\\[7mm] the $e^\pm$ excess from DM annihilations}\\

\medskip
\bigskip\color{black}\vspace{0.6cm}
{
{\large\bf Gianfranco Bertone$^a$, Marco Cirelli$^b$,\\[3mm] Alessandro Strumia$^c$, Marco Taoso$^{a,d}$}
}
\\[7mm]
{\it $^a$ Institut d'Astrophysique de Paris, France. UMR7095-CNRS \\ Universit\'e Pierre et Marie Curie, 98bis Boulevard Arago, 75014 Paris, France} \\
{\it $^b$ Institut de Physique Th\'eorique, CEA-Saclay and CNRS, France\footnote{CEA, DSM, Institut de Physique Th\`eorique, IPhT, CNRS, MPPU, URA2306, Saclay, F-91191 Gif-sur-Yvette, France}} \\
{\it $^c$ Dipartimento di Fisica dell'Universit{\`a} di Pisa and INFN, Italia} \\
{\it $^d$ INFN, Sezione di Padova, via Marzolo 8, Padova, 35131, Italy} 

\bigskip\bigskip\bigskip

{\large
\centerline{\large\bf Abstract}

\begin{quote}
%\large
PAMELA and ATIC recently reported an excess in $e^\pm$ cosmic rays.
We show that if it is due to
Dark Matter annihilations,
the associated gamma-ray flux and the synchrotron emission produced by $e^\pm$ 
in the galactic magnetic field 
 violate HESS and radio observations of 
the galactic center and HESS observations of dwarf Spheroidals, unless the DM 
density profile is significantly less steep than the benchmark NFW and Einasto 
profiles.

\end{quote}}

\end{center}

\newpage

%\tableofcontents

\section{Introduction}

Evidence for the existence of Dark Matter comes from a number of astrophysical and cosmological probes~\cite{reviews}, yet no direct uncontroversial detection has been made so far, and its properties remain largely unknown. 
The excesses in cosmic ray $e^\pm$ spectra suggested by the recent observations of  PAMELA and ATIC-2 in the energy range 10 GeV -- 1 TeV can be tentatively interpreted as due to DM annihilations in the galactic halo. Indeed, the PAMELA satellite reported~\cite{PAMELA} an excess in the positron fraction $(e^+/(e^++e^-))$, with respect to the expected astrophysical background above 10 GeV. The spectrum features a steep rise up to 100 GeV,
the highest energy currently probed by the experiment. No excess is instead seen in the antiproton flux~\cite{PAMELApbar}. The balloon-borne experiment ATIC-2~\cite{ATIC-2} reported the detection of a peak in the undistinguished flux of positron and electrons, in the energy range between 300 and 800 GeV. The two signals are compatible and, at low energy, the precise data from PAMELA confirm previous hints from other experiments~\cite{previous}. 

Of course caution should be used when interpreting the data, as the background flux from conventional astrophysical processes (e.g.\ spallation of cosmic rays on the interstellar gas) carry big uncertainties, as large as one order of magnitude, 
so that even the very evidence for an excess in PAMELA data could possibly be jeopardized~\cite{Salati}. 
A more likely possibility is that the observed effects are due to some single astrophysical object, such as a pulsar (a foreground signal for astrophysics, but an annoying distraction for Dark Matter searches), or a collection of them. Indeed it is expected that pulsars produce a power law spectrum of mostly electron-positron pairs, with a cut-off in the multi-TeV range. The known nearby pulsars Geminga and B0656+14 are the main candidates, but unknown and past pulsars can contribute to the integrated flux~\cite{pulsars}. Work is underway to assess more precisely the features of the expected fluxes from pulsars and discriminate from the DM hypothesis~\cite{pulsars2}.

If however Dark Matter annihilations are at the origin of the observed anomalies in cosmic rays, than the data allow to determine the required properties of the Dark Matter: mass $M$, annihilation cross section $\sigma v$ and main annihilation mode. The analysis in ref.~\cite{CKRS} studied systematically which of these can fit the data, finding three main classes of models.
\begin{itemize}
\item[a)] on the basis of the $e^+$ and $\bar p$
data from PAMELA, the Dark Matter can be: 
\begin{itemize}

\item[a1)] a particle that dominantly annihilates into leptons, with no strong preference for the mass, if above a few hundred GeV;

\item[a2)] a particle that annihilates into $W,Z$ or higgses and that has a mass $\circa{>} 10\TeV$.
An example in this class is the fermion quintuplet in the predictive Minimal Dark Matter model~\cite{MDM1}, where DM has mass $M=9.6\TeV$ and annihilates into transversely polarized $W^+ W^-$: this model predicted the PAMELA excess and predicts that 
the ATIC peak is not there;
\end{itemize}

\item[b)] adding the peak from ATIC-2, a clear indication for the mass emerges: DM has to be a particle with mass $\sim1\TeV$ that dominantly annihilates into leptons. We will exemplify this class of models referring to a generic candidate with $M=1\TeV$ and annihilations into $\mu^+\mu^-$. 

\end{itemize}

The upcoming results of ATIC-4~\cite{ATIC-4}, PAMELA, or the first data from the Fermi LAT calorimeter~\cite{GLASTe+e-} 
or Air Cherenkov Telescopes~\cite{HESSepm}
can soon check if a peak is really present in the $e^+ + e^-$ spectrum just below 1 TeV: if the peak is there b) is favored and a) is excluded; if instead the peak is not there, then a) is favored
and b) excluded. Models with $M \ll 1\TeV$ appear to be already disfavored.

\smallskip

For what concerns the magnitude of the annihilation cross section, the large flux above the background in the PAMELA and ATIC data indicates a very large $\sigma v$ (see fig.~9 of~\cite{CKRS}). For instance, for a candidate in class b), a value of the order of ${\rm few}\, 10^{-23}\cm^3/{\rm sec}$ is needed to fit the data. This is much larger than the typical thermal cross section $\sigma v = 3\cdot 10^{-26}\cm^3/{\rm sec}$ suggested by the cosmological DM abundance. 
%In both examples of type a) and b) we assume $\sigma v = 10^{-23}\cm^3/{\rm sec}$ at $v\sim 10^{-3}$, so that the positron excess can be reproduced with a mild boost factor.
%This cross section is higher than the $\sigma v = 3\cdot 10^{-26}\cm^3/{\rm sec}$ at $v\sim 0.2$ suggested by the cosmological DM abundance:
As discussed in \cite{CKRS,Nima}, the two values can be reconciled if a Sommerfeld enhancement is at work: this effect, in fact, depends on the DM velocity in an important way and so it would be present at $v\sim 10^{-3}$ (the typical velocity of DM particles annihilating in the galactic halo at the present time) and reduced or absent at $v\sim 0.2$ (the velocity at decoupling). More precisely, the enhancement of a non-relativistic $s$-wave DM annihilation~\cite{Somm,MDM2,MDM3,Nima} can be approximatively characterized in terms of two critical velocities $v_{\rm min}$ and $v_{\rm max}$ as follows:
\beq \sigma v = {\rm constant}\times \left\{\begin{array}{ll}
1 & \hbox{for $v>v_{\rm max}$}\\
v_{\rm max}/v & \hbox{for $v_{\rm min}< v <v_{\rm max}$}\\
v_{\rm max}/v_{\rm min} & \hbox{for $v<v_{\rm min}$}\\
\end{array}\right. \quad .\eeq
In terms of particle-physics parameters, assuming that the long-range force that gives rise
to the Sommerfeld enhancement is a vector with mass $M_V \ll M$ and gauge coupling $g_V$ to DM,
one has $v_{\rm max} \approx g_V^2/4$. The value of $v_{\rm min}$ is $v_{\rm min}\approx M_V/M$
unless a (DM DM) bound state with small binding energy $E_B$ is present; in such a case the Sommerfeld effect grows down to a smaller $v_{\rm min} \approx \sqrt{E_B/M}$.
In the exemplar model of class a1), the enhancement is automatically present (via the exchange of weak gauge bosons). Extra states are instead required in class b), so that~\cite{CKRS} suggested that DM might be charged under an extra U(1), proposing a specific model. More proposals along these or different lines have followed~\cite{Nima,post-pamela}.
Alternatively one can invoke either non-thermal DM or very large boost factors \cite{Bringmann}. However, the latter can only arise in rather exotic scenarios (e.g.\ DM mini-spikes around black holes~\cite{Brun:2007tn,Bertone:2005xz}), but not in the framework of DM subhalos with realistic properties~\cite{Lavalle}.

\bigskip

Given these tantalizing but surprising hints of Dark Matter annihilations in the charged particle signals, it is now crucial to consider the constraints on this interpretation that come from the photon fluxes that necessarily accompany such charged particles. These photon fluxes are produced:
\begin{itemize}
\item[i)] directly as a product of the DM annihilations themselves (mainly from the bremsstrahlung of charged particles and the fragmentation of hadrons, e.g. $\pi^0$, produced in the annihilations), at energies comparable to the DM mass $M$, i.e.\ 
in the $\gamma$-ray energy range of tens of GeV to multi-TeV. 

\item[ii)] at much lower energies, e.g.\ radio to visible frequency, by the synchrotron radiation emitted in the galactic magnetic field by the electrons and positrons produced by DM annihilations.
\end{itemize}

The best targets to search for these annihilation signals are regions
with high DM densities, such as the Milky Way Galactic Center (GC), 
the Milky Way Galactic Ridge (GR) and the Sagittarius Dwarf spheroidal 
satellite galaxy (Sgr dSph). The predicted photon fluxes 
can then be compared with observational data, in order to rule out 
combinations of astrophysical and particle physics parameters
that violate observational constraints.

\medskip

The aim of this paper is to compare the regions suggested by the PAMELA (and ATIC) data in the plane of annihilation cross section and DM mass ($\sigma v, M$) with those excluded by photon observations. We perform the analysis for arbitrary values of $M$ and for several different primary annihilation modes. We take into account different choices for the main astrophysical unknown ingredients: the galactic DM density profiles and the galactic magnetic field. 
In section \ref{section:gamma} we discuss bounds from gamma-ray observations, mainly performed by the HESS experiment. Section \ref{section:radio} discusses bounds from lower energy photons radiated by the $e^\pm$.

%\begin{figure}
%\begin{center}
%$$\includegraphics[width=0.99\textwidth]{boundsIsoe}$$
%\caption{\em Same as Fig.\ref{fig:boundsNFW}, but assuming smoother profiles for the Dark Matter distribution: Isothermal core for the Milky Way, 'large cored' for SgrDwarf.
%\label{fig:boundsIso}}
%\end{center}
%\end{figure}

\begin{figure}
\begin{center}
\includegraphics[width=0.45\textwidth]{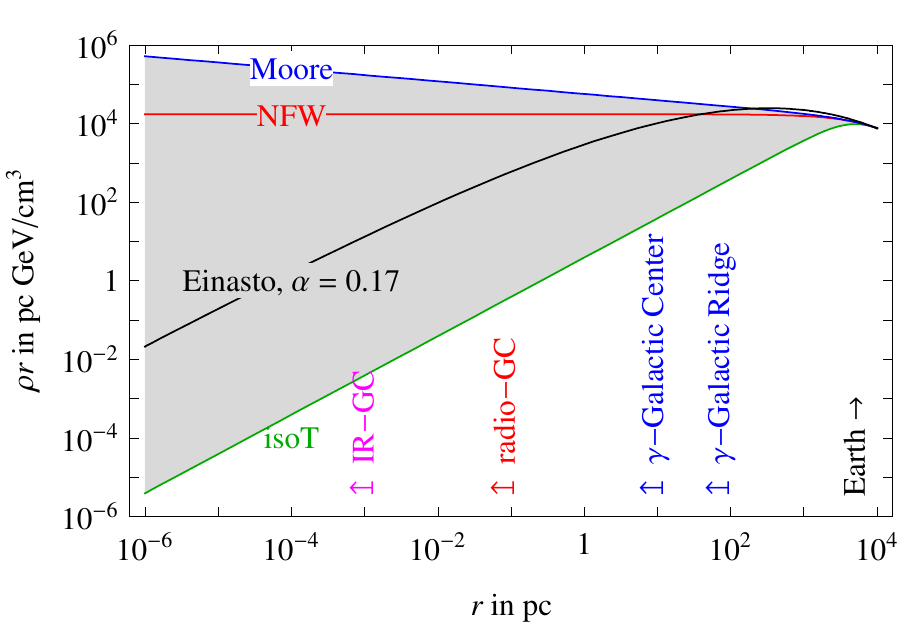}\qquad
\includegraphics[width=0.45\textwidth]{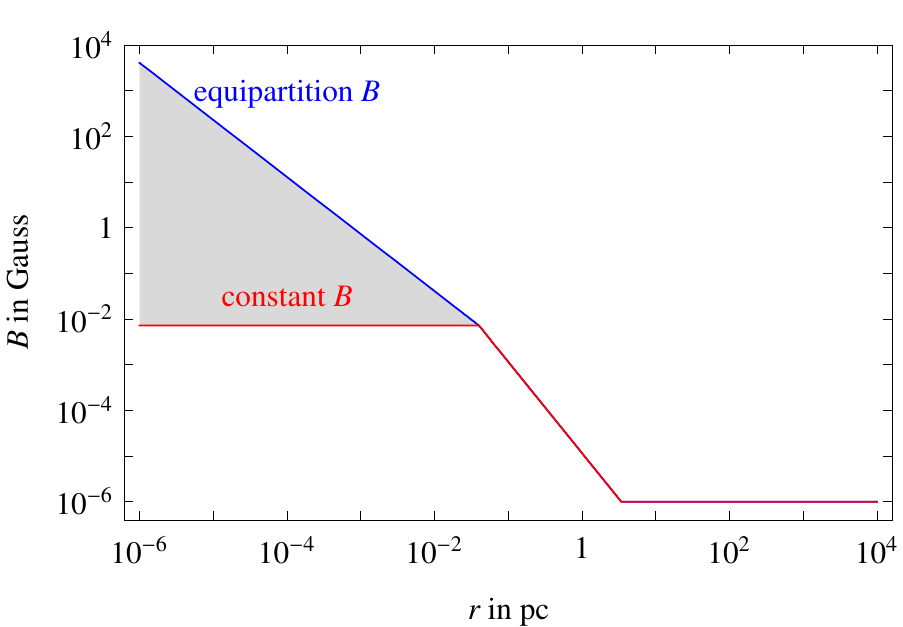}
\caption{\em Shape of DM density (left) and magnetic field (right) profiles discussed in the text, as a function of the galactocentric coordinate $r$.
\label{fig:B}}
\end{center}
\end{figure}

\section{$\gamma$ ray observations}
\label{section:gamma}

We start by considering the $\gamma$-ray fluxes produced by DM annihilations directly. 
Since DM is neutral, a tree-level annihilation into $\gamma$'s is of course not possible, thus the flux is the sum of various effects  that arise at higher order in $\alpha_{\rm em}$: 
i) a continuum at lower energies produced by the bremsstrahlung of charged particles and the fragmentation of hadrons produced in the annihilations; 
ii) a line at $E\approx M$ produced by one-loop effects; 
iii) possibly a continuum at $E$ just below $M$ produced by three-body annihilations~\cite{IB}.
Infrared divergences in the total annihilation rate cancel among i) and one loop corrections
without photons in the final state, and these contributions are separately
gauge invariant in the energy ranges where they are separately relevant. The details of contributions ii) and iii) are model dependent, so that we only consider the contribution i).

The differential flux of photons from a given angular direction $d\Omega$ is
\beq 
\label{gammaflux}
\frac{d \Phi_\gamma}{d\Omega\,dE} = \frac{1}{2}\frac{r_\odot}{4\pi} \frac{\rho_\odot^2}{M_{\rm DM}^2} J
\sum _f\langle \sigma v\rangle_f \frac{dN_\gamma^f}{dE}  ,\qquad
J = \int_{\rm line-of-sight} \frac{ds}{r_\odot} \left(\frac{\rho(r)}{\rho_\odot}\right)^2 
\eeq
where $r_\odot \approx 8.5$ kpc is the distance of the Sun from the galactic center, $\rho_\odot = 0.3\ {\rm GeV/cm}^3$ is the DM density at the location of the solar system and $f$ runs over all the $\gamma$-ray producing channels with annihilation cross section $\langle \sigma v\rangle_f$ and individual spectrum $dN_\gamma^f/dE$. The adimensional quantity $J$ encodes the astrophysical uncertainty. When observing a region with total
angular size $\Delta\Omega$ the factor $J ~d\Omega$ gets replaced by $ \bar J \cdot \Delta\Omega = \int_{\Delta\Omega} J~ d\Omega$.

\begin{center}
\begin{table}
\centering
\begin{tabular}{c|ccc}
MW halo model &$r_s$ in kpc&$\rho_s$ in GeV/cm$^3$&$\bar J\, \left(10^{-5}\right)$ \\
\hline 
NFW~\cite{Navarro:1995iw} & 20 & 0.26 & $15\cdot 10^3$ \\
Einasto~\cite{Einasto} & 20 & 0.06 & $7.6 \cdot 10^3$\\ 
Isothermal~\cite{isothermal} & 5 &1.16 & $13 $\\ 
\end{tabular}
\caption{\em Parameters of the density profiles for the Milky Way discussed in the text and corresponding value of $\bar{J}$ for $\Delta\Omega=10^{-5}$. 
In all cases we imposed the normalization $\rho(r_\odot) =0.3~\GeV/\cm^3$.}
\label{tabprofiles}
\end{table}
\end{center}

In order to compute the flux, one thus has to specify 
the DM density profile as a function of the galactocentric
coordinate, $\rho(r)$. It is commonly assumed that DM 
follows a `universal' Navarro, Frenk and White 
(NFW) profile~\cite{Navarro:1995iw}
\beq\label{eq:NFW}
\rho_{\rm{NFW}}(r) = \frac{\rho_s}{\frac{r}{r_s}\left( 1 + \frac{r}{r_s}\right)^2} \,.
\eeq
%For the Milky Way, we have adopted the parameters in Table ~\ref{tabprofiles}.
It is however unclear whether this analytic formula, obtained by 
fitting DM halos in $N$-body simulations,
actually captures the behaviour of the density profile 
down to the innermost regions. 
In fact, although there is general agreement on the 
shape of profiles at large scales, profiles steeper 
than NFW, with $\rho(r) \propto r^{-1.2}$ at radii much
smaller than the virial radius, have been found to 
provide a better fit to simulated halos \cite{Diemand:2005wv}. 
This claim has been subsequently challenged, e.g.\ by 
Ref.~\cite{Einasto}, where it was found that the
so-called {\it Einasto} profile 
\beq
\rho_{\rm Einasto}(r) = \rho_s\cdot \exp\left[- \frac{2}{\alpha} \left( \left(\frac{r}{r_s}\right)^\alpha -1 \right) \right], 
\qquad \alpha=0.17
\eeq
should be preferred, since the profiles of simulated halos
appeared to become shallower and shallower towards 
the Galactic center, without converging to a definite
power-law. 

Finally, a truncated {\it isothermal} profile
\beq
\rho_{\rm{iso}}(r) = \frac{\rho_s}{1+\left(  \frac{r}{r_s}\right)^2} \,
\eeq
is sometimes adopted as a benchmark, since it is 
representative of `shallow' DM profiles~\cite{isothermal}. 

%Following Ref.~\cite{Navarro:2008kc}, we 
%have adopted here the parameters shown in Table~\ref{tabprofiles};
%note that the value of $\bar{J}\left(\Delta\Omega=10^{-5}\right)$ 
%is in this case only a factor $\sim 2$ smaller than the one for NFW.

In Table~\ref{tabprofiles}, we show the parameters of the 
aforementioned density profiles (plotted in fig.\ref{fig:B}a) for the case of the Milky Way and the value of the 
quantity $J$ averaged over a solid angle $\Delta \Omega = 10^{-5}$
str, corresponding to the angular resolution of gamma-ray
experiments such as HESS and Fermi LAT. Note that, for the Einasto profile, we choose a value of $r_s = 20$ kpc representative of the results of Ref.~\cite{Einasto} for  different simulations; 
the value of $\bar{J}\left(\Delta\Omega=10^{-5}\right)$ is only a factor $\sim 2$ smaller than the one for NFW.

We stress that aside from these uncertainties, the DM distribution is 
further complicated by a number of physical processes that are
not accounted for in most numerical simulations, such as the presence of a 
supermassive black hole that dominates
the gravitational potential within 1 pc from the Galactic
center, and a stellar cusp that inevitably interacts 
with the DM fluid, a circumstance that makes it difficult
to accurately estimate the DM profile in the central 
region (see e.g.\ Ref.~\cite{Bertone:2005hw} and references therein).
We do not include these model-dependent processes in the following. 
More generally, the extrapolation of the numerical Dark Matter profiles in the regions very close to the Galactic Center ($\circa{<} 10$ pc) is of course to be taken with care, as simulations cannot resolve small radii. 
As we will see, however, some of the constraints come from regions as large as ${\cal O}(100)$ pc (the size of the Galactic Ridge region, for instance, or of the Sagittarius Dwarf galaxy) where the impact of these uncertainties is much less important. Keeping in mind these remarks and possible caveats, in the following we will discuss the astrophysical constraints for different choices of the DM profile.

\subsection{$\gamma$-ray observations of the Galactic Center}

\begin{figure}
\begin{center}
$$\includegraphics[width=0.99\textwidth]{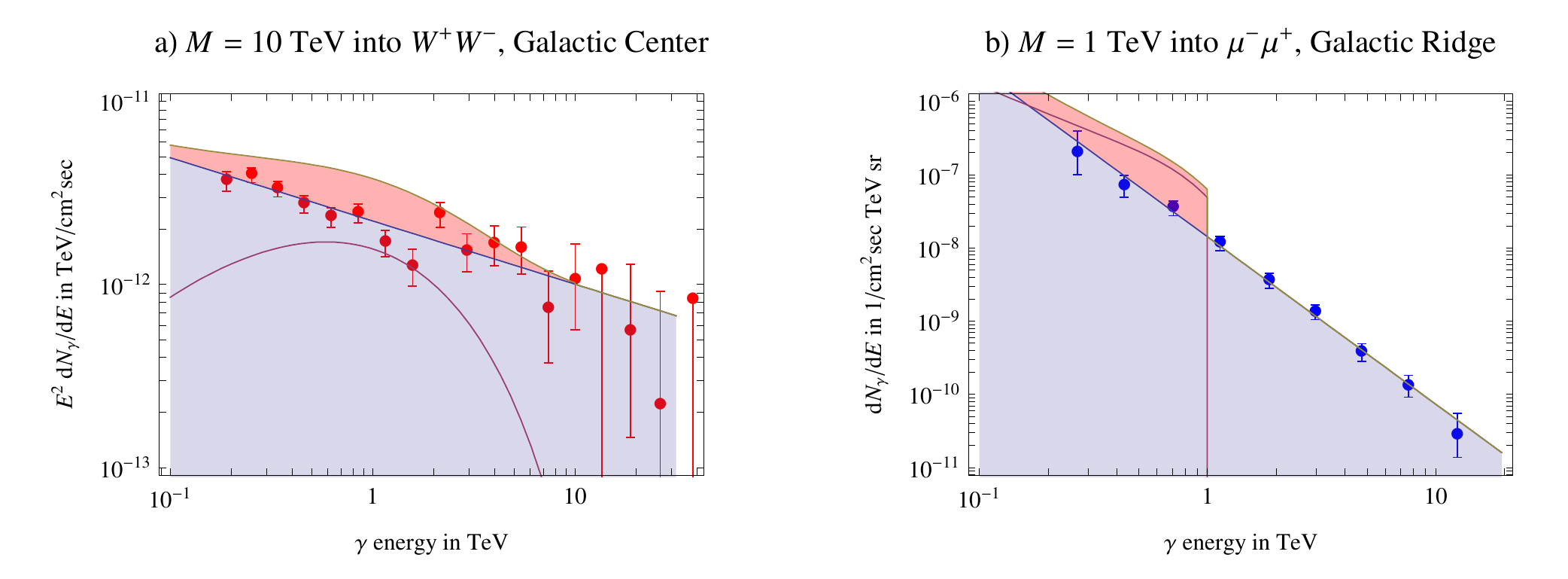}$$
\caption{\em HESS observations of the Galactic Center (left) and Galactic Ridge (right)
compared with the annihilation signals from our two sample models a1) (left) and b) (right), assuming a NFW profile and an annihilation cross section of $\sigma v = 10^{-23}$\, cm$^3$/sec, and with the sum of the annihilation signal and a possible astrophysical background flux. 
\label{fig:SampleHESS}}
\end{center}
\end{figure}

HESS observations in the direction of the Galactic Center have 
revealed a source of Very High Energy $\gamma$-ray emission (HESS J1745-290)
lying within  $7'' \pm 14_{\rm stat}'' \pm 28_{\rm syst}''$ from the 
supermassive black hole Sgr A*, and compatible with a 
point source of size less then $1.2'$~\cite{HessGC}. The corresponding 
energy spectrum, shown in figure \ref{fig:SampleHESS}a, is well fitted 
by a power law $d\Phi_\gamma/dE \propto E^{-2.25\pm0.04}$, over two decades in energy,
and it has been confirmed by the MAGIC collaboration~\cite{Albert:2005kh}.
The EGRET experiment had actually previously reported the detection
of a point source (3EG J1746-2851) within 0.2 degrees from Sgr A*~\cite{mayer}. 
However, a re-analysis based on photons with energies above 1 GeV has 
shown that the source is slightly offset with respect to the galactic 
center~\cite{Hooper:2002ru}.

The possibility to interpret both sets of $\gamma$ observations (separately or
at the same time) in terms
of DM annihilations has been discussed  e.g.\ in 
Refs.~\cite{Cesarini:2003nr,Profumo:2005xd,Bertone:2002ms,Zaharijas:2006qb,Dodelson:2007gd,Regis:2008ij,Bertone:2002je}, 
and Ref.~\cite{Baltz:2008wd} discussed the prospects for the detection of DM with Fermi LAT. Here, we take a conservative 
approach and consider the observed gamma-ray emission as an
upper limit to the DM annihilation flux, in order to test the compatibility with a DM interpretation of the PAMELA data.
We compute the constraints in the $\sigma v$ versus mass plane, by requiring
that DM annihilation flux does not exceed (at $3\sigma$, in terms of the error bars quoted by the HESS collaboration) the observed emission {\it at any data point}.

%According to our conservative fitting procedure, both our sample models are compatible with HESS data
%{\bf (these models can fit the PAMELA excess if a boost factor enhances the $e^+$ DM rate by one or two %orders of magnitude)}
%as illustrated in fig.\fig{SampleHESS}.
%The first model mimics the Minimal Dark Matter theory, where  a full computation of the
%$\gamma$ spectrum is possible and gives a total result that resembles the observed spectrum~\cite{MDM3},
%so that, even if the contribution i) only appears to give an unseen excess in fig.\fig{SampleHESS}b,
%a global fit is possible and the full bound is well approximated by our simplified conservative fitting %procedure.

As an example of our confrontation with data, the left panel of fig.\fig{SampleHESS} shows the gamma-ray flux from the Galactic center (assuming a NFW profile) produced by the annihilations of 10 TeV DM particles into $W^+ W^-$, the aforementioned sample model a1), with an annihilation cross section $\sigma v = 10^{-23}\ {\rm cm}^3/{\rm sec}$. This mimics the Minimal Dark Matter theory~\cite{MDM3}. As one can see, the DM $\gamma$ flux does not exceed any of the HESS data points, and correspondingly this point in parameter space will lie in the allowed region of fig.\fig{boundsNFWq} (see below). The left panel of fig.\fig{SampleHESS} also shows the superposition of the DM signal with a sample power-law background: while in this case one would conclude that the model is excluded because the summed flux exceeds the HESS observations by more than $3\sigma$ at several data points, this conclusion would not be solid. Indeed, choosing a different background (e.g.\ lower in normalization) within its large uncertainties can re-allow the model. Adopting the criterion of comparing each single data point with the DM-only flux allows us to have more conservative and robust results, in the sense that they can only be made tighter by specific choices for the background, that we do not perform in order not to exclude more models than it is justified by current data.
Moreover, we recall that the signal from DM annihilations that we consider here does not include all the model dependent contributions (see the discussion at the beginning of sec.~\ref{section:gamma}). 
A full computation of the gamma-ray spectrum and of its uncertainties to be inserted in a global fit
is possible only in specific theories.
For instance for the case of Minimal Dark Matter it has been obtained in Ref.~\cite{MDM3}. These contributions can change somewhat the shape of the DM signal and bring it closer to the shape of the observed spectrum, even in absence of a power-law background. For this additional reason, it is apparent that it would be wrong to exclude a model such as the one illustrated in the left panel of fig.\fig{SampleHESS}.

%\begin{figure}
%\begin{center}
%$$\includegraphics[width=0.99\textwidth]{SampleHESSRidge}$$
%\caption{\em HESS observations of the Galactic Ridge region compared with two models suggested by the PAMELA excess, assuming a NFW profile.
%\label{fig:SampleHESSRidge}}
%\end{center}
%\end{figure}

\bigskip

The HESS collaboration has also recently discovered a diffuse gamma-ray emission, correlated spatially with the Galactic Ridge (GR), a complex of giant molecular clouds in the central 200 pc of the Milky Way~\cite{HessRidge}. Once point sources, including HESS J1745-290, are subtracted, the reconstructed 
gamma-ray spectrum for the region with galactic longitude
$-0.8^\circ<\ell < 0.8^\circ$ and latitude $|b| < 0.3^\circ$ is well described by a power
law with photon index $\Gamma = 2.29 \pm 0.07_{\rm stat} \pm 0.20_{\rm syst}$. In this region,
the predicted DM signal is smaller than in a small cone pointing towards the
Galactic center, but the astrophysical background is also significantly
reduced, and the constraints are less sensitive to the slope of the DM density
profile. 
Fig.\fig{SampleHESS}b shows the HESS\ data and the signal in our sample model b), i.e.\ a candidate with $M=1$ TeV annihilating to $\mu^+\mu^-$, assuming a NFW profile. The cross section has been taken here to be  $\sigma v = 10^{-23}\ {\rm cm}^3/{\rm sec}$ as in the left panel. 
The same discussion as above applies: what is seen here is that the DM $\gamma$ signal exceeds a couple of data points by more than 3$\sigma$ and therefore the model will lie within the excluded region.

\bigskip

In figures\fig{boundsNFW} and\fig{boundsNFWq} we show the results of the analysis of the data described above.
The continuous blue lines shows our conservative bounds on the annihilation cross section $\sigma v$ from HESS\ observations of the Galactic Center, and the dot-dashed blue lines show the comparable bounds from Galactic Ridge observations. Figs.\fig{boundsNFW} refer to DM annihilation into leptons, while fig.s\fig{boundsNFWq} show the more `traditional' DM annihilation modes into $W^+W^-$, $b\bar{b}$ and $t\bar{t}$. Barring the possibility of boost factors or Sommerfeld enhancements different for $\gamma$ and $e^\pm$ observations, we see in fig.s\fig{boundsNFW} that the green regions that can fit the PAMELA anomaly (and the red regions that can also fit the ATIC anomaly) are excluded for masses $M\circa{>}300\GeV$, by two orders of magnitude if DM follows the NFW density profile, by an order unity factor if DM follows the Einasto profile, and are allowed if DM follows the isothermal profile (somewhat disfavored, however, by $N$-body simulations).\\ 
In fig.s\fig{boundsNFWq} a similar situation holds. The green PAMELA bands are here truncated because low DM masses do not allow a good fit to the anti-proton data; the truncation is conservatively put at 1 TeV, but masses up to multi-TeV still do not give a good fit, as discussed in the Introduction (see~\cite{CKRS} for the full analysis). 

%%%%%%

\subsection{Sagittarius dwarf spheroidal galaxy}
\label{section:SgrDwarf}

Dwarf spheroidal galaxies are among the most DM-dominated structures, so that they allow to search for $\gamma$ ray signals of DM annihilations with minimal astrophysical backgrounds. In particular, HESS has observed the Sagittarius Dwarf galaxy~\cite{HessSgrDwarf}, a satellite of the Milky Way which is located at a distance of $d=24\,{\rm kpc}$ from the Sun. The satellite is thought to be in the process of being disrupted by multiple passages through the Milky Way disk, and the fact that it still exists is taken as an indication of the existence of a substantial amount of Dark Matter in it. 

The DM density profile in Dwarf Galaxies is uncertain as much as the one in the Milky Way, with which it might have some correlations. For Sgr dSph we consider the possibilities of a cusped NFW profile~\cite{HessSgrDwarf,SarkarDwarf}  with
density given by eq.\eq{NFW} 
and of the class of cored profiles
\beq
\rho_{\rm core}(r) = \frac{v_a^2}{4\pi G_N} \frac{3r_c^2+r^2}{(r_c^2+r^2)^2}.
\eeq
The normalization factors and the characteristic radii are reported in Table~\ref{tabprofilesDwarf}, where also the corresponding values of $\bar J$, defined according to eq.~(\ref{gammaflux}), are given.
% and $\bar{\tilde J} = r_\odot \rho_\odot^2  J$ are given. 
The area of observation corresponds to an aperture angle of $0.14^\circ$ i.e.\ to a size of $\Delta \Omega = 2 \cdot 10^{-5}$~\cite{HessSgrDwarf}.

\begin{table}
\begin{tabular}{c|cccc}

Sgr dSph halo & Parameters & Core/Scale radius   & $\bar J\, (2 \cdot 10^{-5})$ & $\bar {\mathcal J}\, (2 \cdot  10^{-5})$\\
\hline
Small core~\cite{HessSgrDwarf} 	& 	$v_a$ = 13.4 km/s  & $r_c$ = 1.5 pc  & $ 31 \cdot 10^3$ & $74\ 10^{24}\,{\rm GeV}^2/{\rm cm}^5 $ \\
NFW~\cite{NFW} 	&	$\rho_s$ = $5.2\GeV/{\rm cm}^3$       &     $r_s$ = 0.62 kpc   &   $1 \cdot 10^3$ &  $2.46\ 10^{24}\,{\rm GeV}^2/{\rm cm}^5 $    \\
Large core~\cite{SarkarDwarf} 		& 	 $v_a$ = 22.9 km/s	& $r_c$ = 0.23 kpc & $0.14 \cdot 10^3$ & $0.32\ 10^{24}\,{\rm GeV}^2/{\rm cm}^5 $ \\
\end{tabular}
\caption{\em Parameters of the density profiles for the Sagittarius Dwarf galaxy discussed in the text and the corresponding value of $\bar{J}(\Delta \Omega)$ (normalized by convention in terms of the solar quantities $r_\odot$ and $\rho_\odot$, as in eq.~(\ref{gammaflux})) for $\Delta\Omega=2 \cdot 10^{-5}$. For reference, the value of the rescaled $\bar{\mathcal J}(\Delta \Omega) = r_\odot \rho_\odot^2 \, \bar{J}(\Delta \Omega)$ is also given.} 
\label{tabprofilesDwarf}
\end{table}

\bigskip

HESS has observed Sagittarius Dwarf for $T_{\rm obs}=11\,{\rm h}$ finding no $\gamma$-ray excess: the integrated photon flux is $N_\gamma \circa{<}85$ at $3\sigma$ ($N_\gamma < 56$ at 95\% CL~\cite{HessSgrDwarf}). Hence an upper bound can be imposed on the annihilation cross section 
\beq 
\sigma v < \frac{8\pi}{T_{\rm obs}} \frac{M^2}{r_\odot \rho_\odot^2 \bar J \Delta\Omega} \frac{N_\gamma}{\displaystyle{ \int dE~A_{\rm eff}(E)\, \frac{dN_\gamma}{dE}}}
\eeq
where the effective area of HESS for observations at $\sim 20^\circ$ (Sgr dSph is located at $14^\circ$ galactic latitude) $A_{\rm eff}(E)\sim 10^5\,{\rm m}^2$ in the range $E \circa{>} 70$ GeV is taken from~\cite{AeffHESS}.

\bigskip

The resulting bounds  on $\sigma v$ are shown as dashed blue lines in figures\fig{boundsNFW} and\fig{boundsNFWq}. The top rows of the figures assume a NFW DM density profile in Sgr dSph: the bounds are overall comparable or slightly less powerful than the bounds from the Galactic Center and Ridge. In all the lower rows we use  for Sgr dSph a `large core' profile, which gives the minimum $\gamma$ flux among the profiles considered in the literature. The bound becomes the most constraining one when the Milky Way profile is taken to be isothermal. We have not explored whether even smoother profiles of Sgr dSph can be designed (compatibly with observations) that can lift such bound.
It is interesting to note that the typical velocity dispersion of DM in Dwarf Spheroidal galaxies is about $10\,{\rm km/s}$~\cite{SarkarDwarf}, smaller than in our galaxy:
thereby their constraint becomes stronger and dominant for
models where light particles give a Sommerfeld enhancement down to a small $v_{\rm min}$~\cite{Nima}.

\bigskip 

Notice that the regions suggested by PAMELA for light DM mass are not probed by HESS\ observations due to its high energy threshold. DM  that annihilates into leptons tends to give most of the signal at $\gamma$ energies just below the DM mass: the forthcoming Fermi/GLAST $\gamma$ observations are not expected to be very significant for our purposes, as they will extend the HESS\ observations down to lower energies but will overlap
with the HESS\ observation at the $\circa{>}100\GeV$ gamma energies suggested by the PAMELA/ATIC excesses.
We now turn to radio-wave observations.

\section{Radio observations of the Galactic Center}\label{section:radio}
The $e^\pm$ produced by DM annihilations within the galactic magnetic field radiate synchrotron radiation.
The Galactic Center is presumably the best region to search for this effect, because
of the large local value of the  DM density and magnetic fields~\cite{Gondolo:2000pn,Bertone:2002je,Bertone:2002ms,Bertone:2001jv,Aloisio:2004hy,Bergstrom:2006ny,Regis:2008ij}. We first detail the necessary astrophysical and particle physics ingredients, and then move to the comparison with observations.

The GC region contains a black hole with mass $M_{\rm BH} \approx 4.3\times 10^6 M_\odot$ (see e.g.
the recent Ref.~\cite{Gillessen:2008qv} and references therein).
This implies two length-scales: the Schwarzschild radius
$R_{\rm BH} =2G_N M_{\rm BH}\approx 4 \times 10^{-7}\,{\rm pc}$  
and the radius of the accretion region, $R_{\rm acc} \equiv 0.04\,{\rm pc}$,
defined to be the region where the velocity flow due to the gravity of the black hole, $v = - \sqrt{R_{\rm BH}/r}$
is larger than the  random galactic motion, $v\sim 10^{-3}$.

Assuming a constant accretion of the BH mass, $\dot M_{\rm BH}\approx 5\cdot 10^{-12}~M_\odot/{\rm sec}$, the matter density is given by
$\rho(r) = \dot M_{\rm BH}/4\pi r^2 v(r) \propto r^{-3/2}$.
Assuming  equipartition of the matter kinetic energy with the magnetic pressure, $\rho v^2/2 = B^2/2$,
the magnetic field is $B(r<R_{\rm acc}) = \sqrt{\rho v^2} = 7.2\,{\rm mG} \cdot (R_{\rm acc}/r)^{5/4}$.
Outside the accretion region, assuming that the magnetic flux is conserved, $B(r>R_{\rm acc})$ scales as $1/r^2$, down to the typical constant galactic value
$B\sim \mu{\rm G}$ reached at $r\sim 100\, R_{\rm acc}$.
This defines the `equipartition' magnetic field, plotted in fig.\fig{B}b.
Another possibility is that $B$ stays constant inside the accretion region: this defines the `constant' magnetic field, again plotted in fig.\fig{B}b.
As we will see, bounds from observations at lowest frequencies are robust, and mildly vary even when
the magnetic field is varied within these extremal possibilities.

Next, we need to compute the number density $n_e (r,p,t)$
of the $e^\pm$ generated by DM annihilations.
We assume stationary conditions, spherical symmetry,
and, in view of the large magnetic fields, we neglect diffusion and assume that
synchrotron radiation dominates energy losses.
%\footnote{Although the GC region has a bipolar structure, such a `spherical-cow' approximation should be sufficient for our purposes.}
Writing the injection term (numerical coefficients are given in particle-physics natural units in the following) 
$Q= \sigma v (\rho^2/2M^2)(dN_{e^\pm}/dE)$ as
$Q = 4\pi p^2 q$ and the number density as 
$n_e = 4\pi p^2 f$, $f(r,p)$ obeys the equation
\beq\label{eq:GCevo}
v \frac{\partial f}{\partial r} + \frac{1}{p^2} \frac{\partial }{\partial p}\left[\dot p_{\rm syn} p^2 f\right]
+ \dot p_{\rm adv}   \frac{\partial f}{\partial p}=q\eeq 
with energy losses
\beq \dot p_{\rm adv} = -\frac{p}{3r^2} \frac{\partial( r^2 v)}{\partial r},\qquad
\dot p_{\rm syn} =
\frac{e^4 B^2 Ep}{9\pi m_e^4}.\eeq
We assume $p\gg m_e$, such that $E\simeq p$ and $\dot E_{\rm syn}\simeq \dot p_{\rm syn}$.
If one can neglect advection (because $\dot p_{\rm adv}\ll \dot p_{\rm syn}$: this happens at
$r>R_{\rm acc}$ and at large $p$), the above equation is solved as
\beq\label{eq:nenoadv}
n_e (r,E) \simeq  \frac{1}{\dot E_{\rm syn}} \int_E^\infty dE'~Q(E',r) =\sigma v \frac{\rho^2}{2M^2}\frac{N_e(E)}{\dot E_{\rm syn}}\eeq
where $N_e(E)$ is the number of $e^+$ or $e^-$ generated with energy larger than $E$ in one DM annihilation.

\medskip

Finally, we can now compute the synchrotron  power $W_{\rm syn}$ generated by the $n_e$ electrons and positrons
in the turbulent magnetic field $B$
\beq\label{eq:syn}
\frac{dW_{\rm syn}}{d\nu}=\frac{\sqrt{3}}{6\pi}\frac{e^3B}{m_e} F(\frac{\nu}{\nu_{\rm syn}}),\qquad
F(x)=x \int_x^\infty K_{5/3}(\xi)d\xi  \approx \frac{8\pi}{9\sqrt{3}} \delta(x-1/3)\eeq
where 
\beq\label{eq:nusyn}
\nu_{\rm syn} = 
\frac{3eB p^2}{4\pi m_e^3}=
4.2\,{\rm MHz} \frac{B}{\rm G} \left(\frac{p}{m_e}\right)^2.\eeq
Reducing the  magnetic field $B$,  the spectrum of synchrotron radiation moves to lower energies,
but the total energy into synchrotron radiation remains constant, until $B$ becomes
so small that other energy-loss mechanisms start to dominate.

Inserting eq.\eq{nenoadv} in\eq{syn} we find
\beq \nu \frac{dW_{\rm syn}}{d\nu} =\frac{\sigma v}{2M^2}\int_{\rm cone} dV~\rho^2~p~ N_e(p) \label{synflux} \eeq
where the integral extends over the observed volume
and $p$ is obtained from eq.s\eq{syn} and\eq{nusyn} as $p=\sqrt{4\pi m_e^3\nu/B}=0.43\GeV (\nu/{\rm GHz})^{1/2}(B/{\rm mG})^{-1/2}$.
A lower $B$ leads to a higher synchrotron flux at the low frequency we consider.

\bigskip

\begin{figure}[p]
\begin{center}
$$\includegraphics[width=0.99\textwidth]{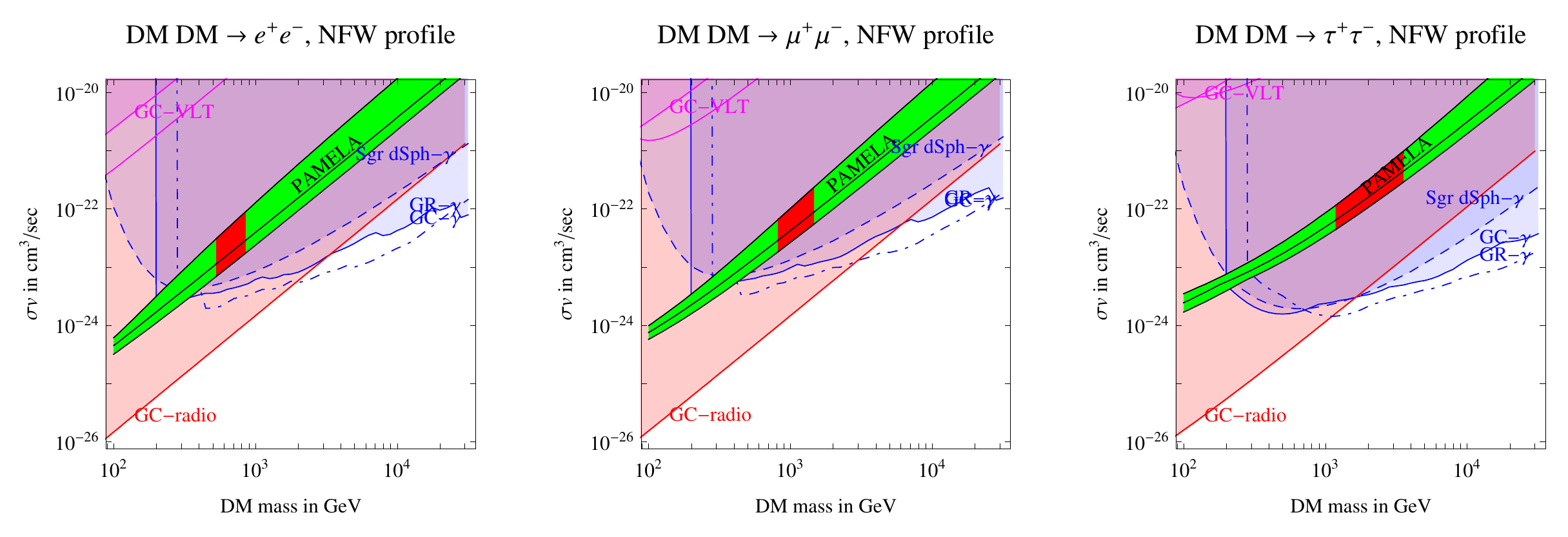}$$
$$\includegraphics[width=0.99\textwidth]{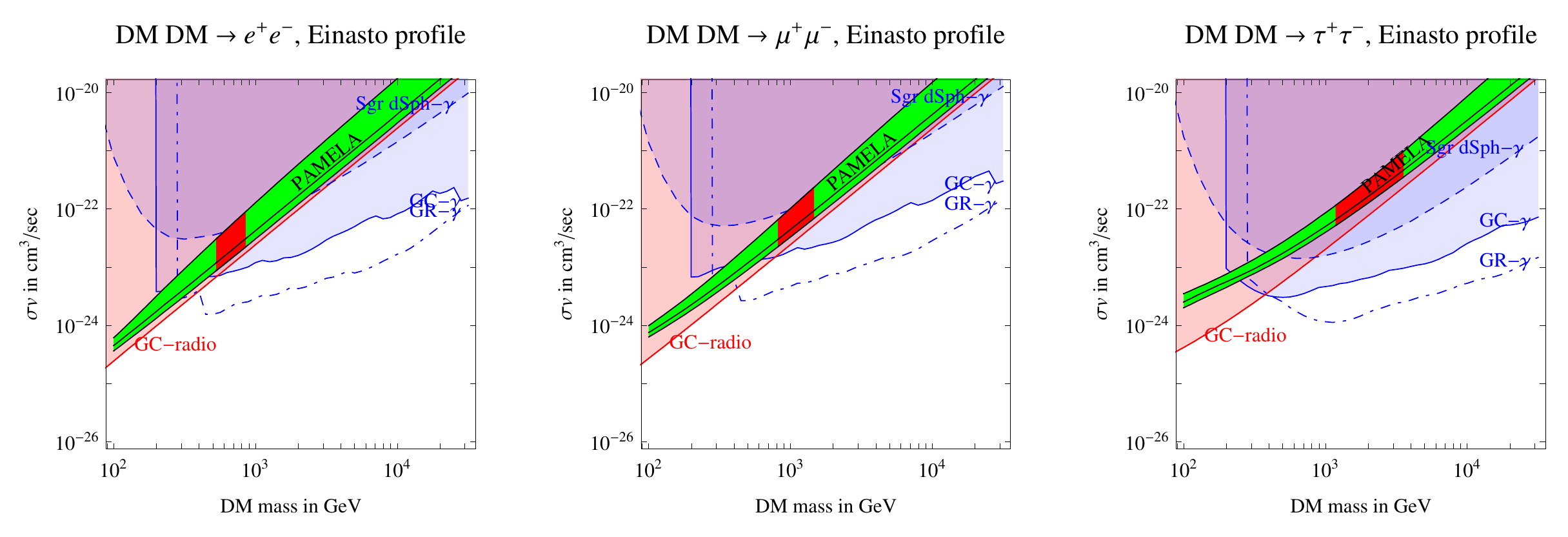}$$
$$\includegraphics[width=0.99\textwidth]{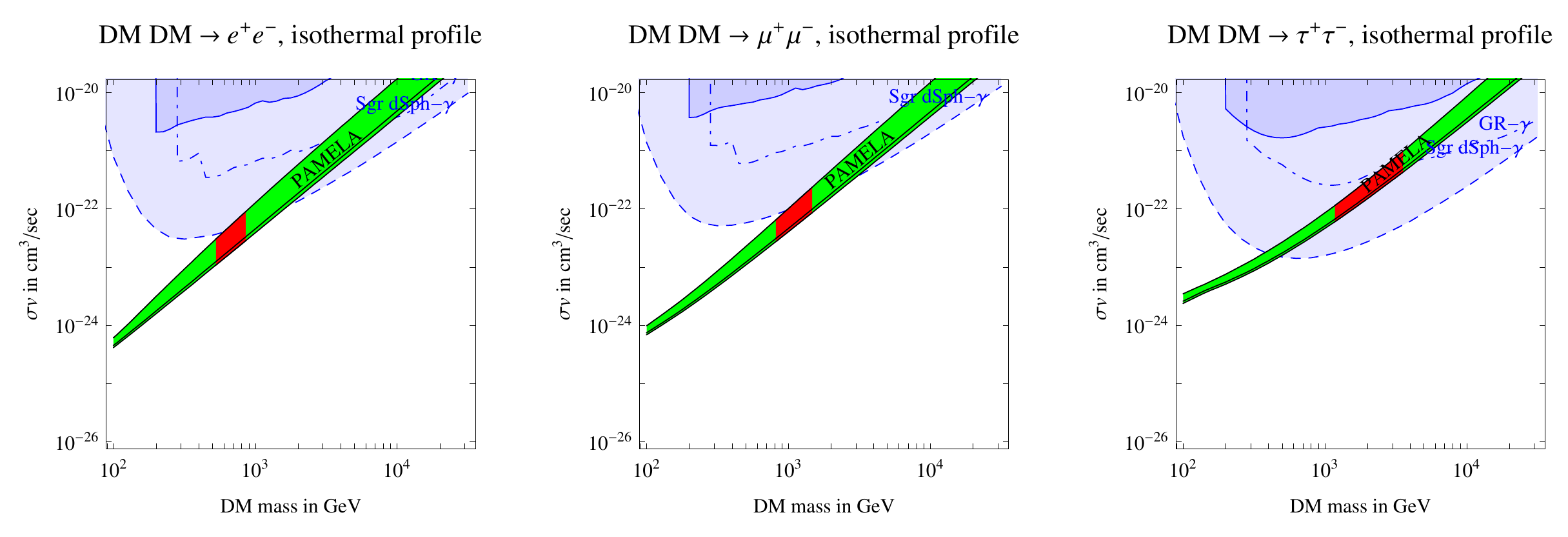}$$
\caption{\em We compare the region favored by PAMELA (green bands)
and ATIC (red regions within the bands)
with the bounds from HESS observations of the Galatic Center~\cite{HessGC}
(blue continuous line), Galactic Ridge~\cite{HessRidge} (blue dot-dashed),
and SgrDwarf~\cite{HessSgrDwarf} (blue dashed) 
and of observations of the Galactic Center at radio-frequencies $\nu=408\,{\rm GHz}$  by Davies et al.~\cite{Davies} (red lines) and at $\nu \sim 10^{14}\,{\rm Hz}$ by VLT~\cite{VLT} (upper purple lines, when present, for equipartition and constant magnetic field).
We considered DM annihilations into $e^+ e^-$ (left column), $\mu^+\mu^-$ (middle), $\tau^+\tau^-$ (right),
unity boost and Sommerfeld factors and
the NFW (upper row), Einasto (middle), isothermal (lower) MW DM density profiles and 
the NFW (upper), large core (middle and lower) Sgr dSph DM density profiles.
\label{fig:boundsNFW}}
\end{center}
\end{figure}

\begin{figure}[p]
\begin{center}
$$\includegraphics[width=0.99\textwidth]{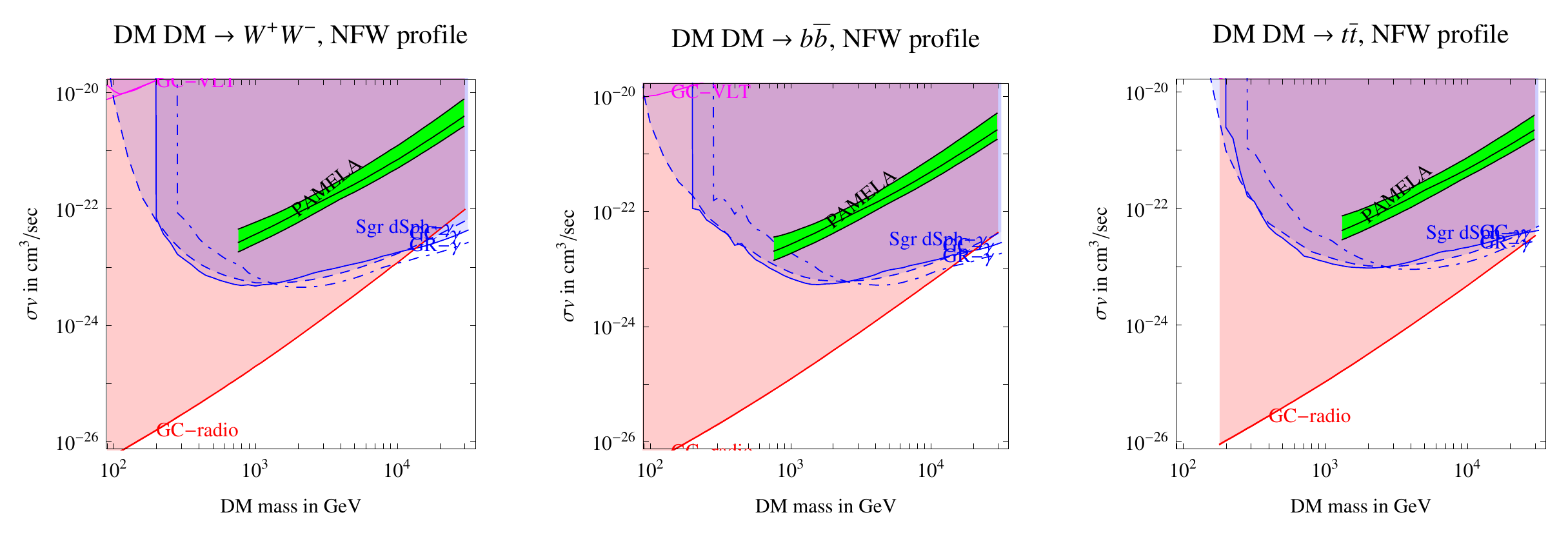}$$
$$\includegraphics[width=0.99\textwidth]{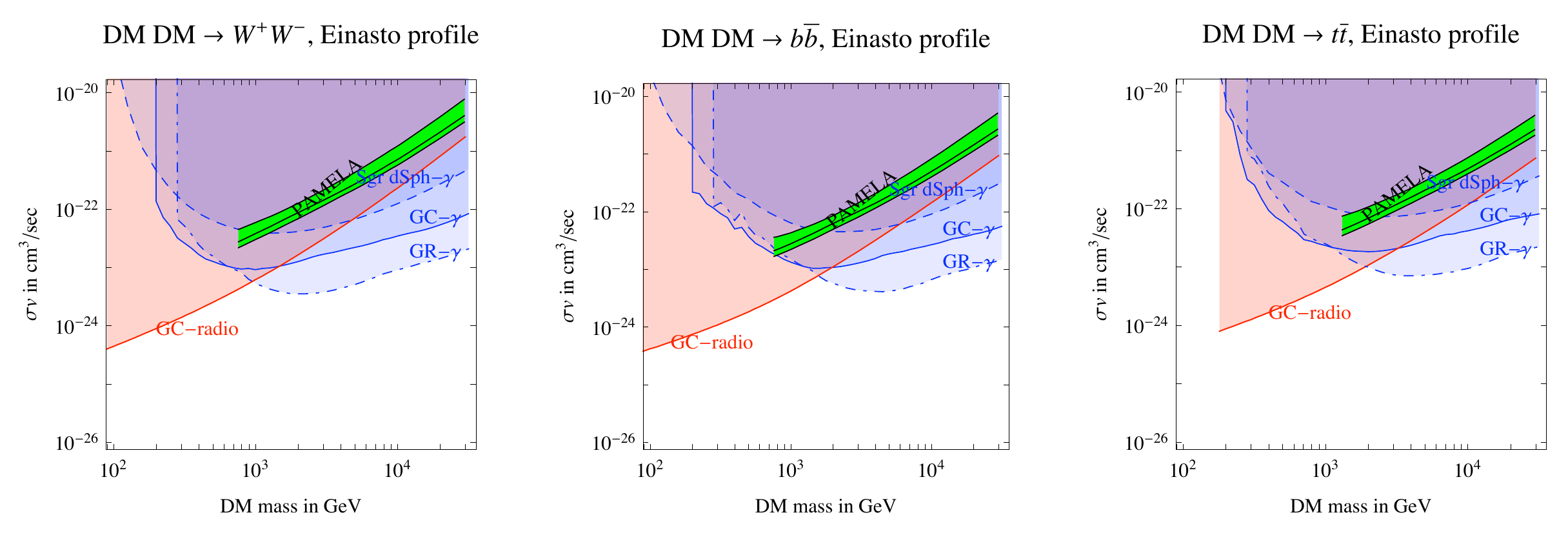}$$
$$\includegraphics[width=0.99\textwidth]{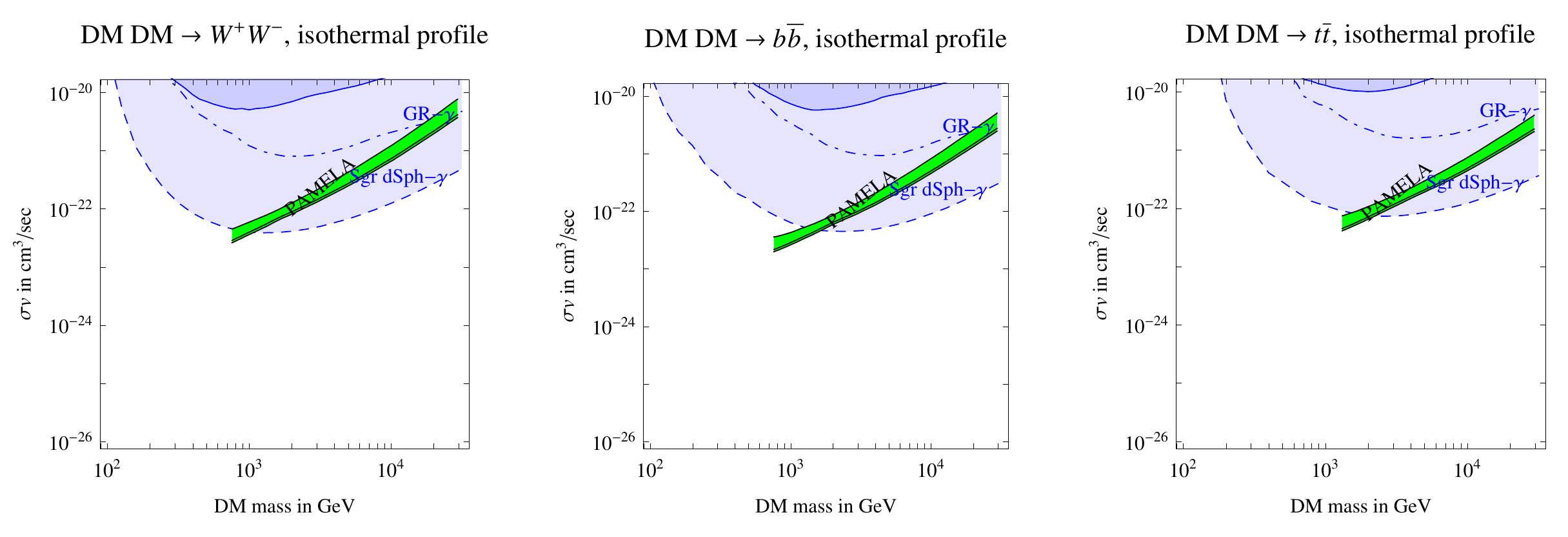}$$
\caption{\em As in the previous fig.\fig{boundsNFW}, but for the cases of 
DM annihilations into $W^+ W^-$ (left), $b\bar b$ (middle), $t\bar t$ (right).
\label{fig:boundsNFWq}}
\end{center}
\end{figure}

We now move to the comparison with observations. 
Since the  observed GC microwave spectrum is harder than what DM decays can produce,
the dominant bound is obtained considering the observation available at the lowest observed
frequency, $\nu=0.408\,{\rm GHz}$, performed by~\cite{Davies} in a region with full width half maximum of $4''$. The observation found an upper limit to the measured flux $S = (\nu \, dW_{\rm syn}/d\nu)/(4\pi r_\odot^2)
 < 2~10^{-16}\,{\rm erg}/{\rm cm}^2{\rm sec}$, that constraints from above the flux in eq.\ref{synflux}. The resulting bounds are plotted in fig.s\fig{boundsNFW} and\fig{boundsNFWq} as red lines. What is seen is that this constraint excludes a large portion of the parameter space for NFW and Einasto DM profiles. The constraint extends to low DM masses (where the $\gamma$-ray bounds from HESS are not effective).
The variation of the magnetic field between `equipartion' and `constant' in the inner region at $r< R_{\rm acc}$
negligibly affects the bound, because the radio emission is predominantly produced by the outer region.
We have also verified that for the relatively shallow profiles under consideration, the synchrotron self-absorption is negligible. 

\medskip

The subdominant  bound (purple lines) comes from the VLT observation at the 
larger infrared/visible frequency, $\nu=0.5~10^5\,{\rm GHz}$:
$S < 3~10^{-12}\,{\rm erg}/{\rm cm}^2{\rm sec}$
from a region with angular size $0.04''$ i.e.\ $r<0.0016\,{\rm pc}$.
It somewhat depends on the magnetic field profile, and
it becomes numerically significant only for spiked DM density profiles~\cite{Regis:2008ij}.
Similarly, observations at higher frequencies give possibly strong but not robust bounds
~\cite{Regis:2008ij},
that also strongly depend on the possibility of having an intense `equipartition' magnetic field close to the Milky Way black hole.

\bigskip

Finally, we need to consider the possible effect of advection, that we neglected so far.
We numerically studied it, and, as it depends on the density profile, magnetic field, DM mass and annihilation mode
we try to qualitatively summarize some general lessons rather than presenting
a large number of specific plots.
At VLT-like (and higher) frequencies, $\nu \sim 10^{14}\,{\rm Hz}$,
synchrotron radiation is dominantly generated by
$e^\pm$ with energy above 1 GeV. 
The density of such higher energy $e^\pm$ is negligibly affected by advection.
Presumably, advection can be neglected also for  the observation by~\cite{Davies} at the lower frequency $\nu=0.408$~GHz,
because the angular aperture of $4''$ corresponds to a region with size
$R_{\rm obs} =0.14$ pc, which is larger than the (presumed) accretion radius $R_{\rm acc}\approx 0.04$ pc,
inside which advection is a significant effect\footnote{Advection typically reduces the $e^\pm$ density  just below $R_{\rm acc}$
and $e^\pm$ accumulate in the inner region.}.  
Advection can be neglected because the total luminosity is dominated by radii comparable to $R_{\rm obs}$,
unless one considers DM density profiles that grow at $r \to 0$ more strongly than the NFW profile,
leading to very strong constraints possibly affected by advection
(and by the reduction of the Sommerfeld enhancement, as the DM velocity grows, becoming relativistic
around the Schwarzschild radius).
The Davies bounds can possibly be weakened if the accretion region is larger.

\section{Conclusions}
We explored the compatibility of the interpretation in terms of DM annihilations
of the excesses in the CR $e^\pm$ spectra claimed by PAMELA and ATIC
with observations of photons at gamma and radio frequencies,
inevitably produced by brehmstrahlung (in DM annihilations) and synchrotron radiation (of $e^\pm$ in galactic magnetic fields) respectively.

\medskip

Figures \ref{fig:boundsNFW} and\fig{boundsNFWq} summarize our results.
The green (red) bands show the region allowed by PAMELA (PAMELA and ATIC combined),
varying the  $e^\pm$ propagation models between the so-called min/med/max sets of parameters (see e.g.~\cite{MDM3, minmedmax} and references therein).
We do not show the subleading $\circa{<}\pm 20\%$ experimental and background uncertainty.
The regions shaded in blue (red) are excluded by our fit of gamma (radio) observations.
One sees that the two data-sets are incompatible if the DM density profile is Einasto, NFW or steeper:
the gamma and radio observations are violated by one or two orders of magnitude. As we conservatively fitted the various data-sets, our results are robust.

\medskip

Plots are made assuming unit boost factors and Sommerfeld enhancement,
and our results remain unchanged (up to a rescaling) if these factors are constant.
In line of principle, it is possible that the $e^\pm$ excess (if dominantly due to DM annihilations in the solar neighborhood)
is significantly more enhanced than DM annihilations around the Galactic Center.
However, numerical simulations suggest that boost factors much larger than unity are
unlikely~\cite{Lavalle}, and we have verified that a realistic subhalo 
population among those discussed in Ref.~\cite{Pieri:2007ir} actually produces 
$\cal O$(1) boost factors for $\sim100$ GeV positrons and gamma-rays in 
a $10^{-5}$ cone towards the galactic center. It should however be mentioned that
a boost factor of a few units is possible for positrons, and the first effect of adding 
substructures to the the smooth DM component of the Milky Way actually is to
{\it reduce} the flux from the Galactic center by a factor $(1-f)^2$, where $f$ is the 
fraction of the mass 
of the smooth halo that goes into clumps, which is expected to be $\cal O$(0.1). 
In absence of a precise prescription, we limit ourselves
to caution the reader that this introduces an $\cal O$(1--10) uncertainty on the 
exclusion plots discussed above.
Similarly, the variation in the Sommerfeld enhancement due to the different 
DM velocity dispersion at the Galactic center can only lead to $\cal O$(1)
uncertainties on the constraints, since the annihilation signal does not 
arise from regions near the Galactic Black Hole horizon where DM has a larger velocity dispersion.

\medskip

In order to perform a model-independent analysis, we considered DM annihilations into pairs of SM particles.
Recently, it was proposed that DM might instead annihilate into some new {\em light}
particle with mass $m\circa{<}m_p$ that can only decay
into SM leptons $\ell$ or pions in view of kinematical constraints~\cite{Nima}.
Such models lead to a $\gamma$-ray flux reduced by about a factor $\ln (M/m_\ell)/\ln (m/m_\ell) \sim 2$,
only mildly alleviating the HESS bounds considered in this paper.
% which is not enough to circumvent the HESS\  bounds considered in this paper.

The DM annihilation interpretation of the PAMELA/ATIC excesses is compatible with gamma and radio observations
if the DM density profile is significantly less steep than the Einasto and NFW profiles.
We also notice that this would mean that $e^\pm$ observed by PAMELA and ATIC dominantly come from regions of the galaxy that are close to us (as the isothermal profile predicts less concentration of DM at the galactic center) and thereby suffer little energy losses. This implies in particular that the direct annihilation channel $\DM\ \DM\to e^+ e^-$ is disfavored: in this case the spectrum would remain close to a peak at $E= M$, which seems to be disfavored by the PAMELA+ATIC $e^\pm$ spectra (that show a broader shape). In other words, a combined fit of PAMELA, ATIC and photon data from a $e^+e^-$ primary channel does not yield a good fit for smooth DM density profiles. Other channels that produce broader spectra (such as $\mu^+\mu^-$) do a better job.

Our results are compatible with other recent analyses~\cite{Bell}, including those~\cite{Regis:2008ij} performed assuming the very steep DM density profiles of Ref.~\cite{Bertone:2005hw}.

\paragraph{Acknowledgements} 
We thank Dario Grasso, Emmanuel Moulin, Lidia Pieri and Marco Regis for very useful discussions. M.C. thanks the Instituto de Fisica Teorica of the State University of S\~ao Paulo, Brazil, for hospitality during the completion of this work. We thank the EU Marie Curie Research \& Training network ``UniverseNet" (MRTN-CT-2006-035863) for support.

\footnotesize

\begin{multicols}{2}

\end{multicols}

\end{document}